# Irreversibility Field and Upper Critical Field in YBa$_2$Cu$_3$O$_{7-\delta}$ thin film

H. El Hamidi[1], A. Taoufik[1], B. El Moudden[1], A. Hafid[1], F. Chibane[1], A. Labrag[1], M. Bghour[1], A. Tirbiyine[1], Y. Ait ahmed[1].

[1] Laboratoire des Matériaux Supraconducteurs, University of Ibn Zohr, Faculté des Sciences, BP: 8106, Agadir, Morocco.

## Abstract

We present a detailed study of the electrical transport properties of YBa$_2$Cu$_3$O$_{7-\delta}$ thin film. The irreversibility fields ($\mu_0 H_{irr}$), upper critical fields ($\mu_0 H_{C2}$), penetration depths ($\lambda$) and coherence lengths ($\xi$) of the YBa$_2$Cu$_3$O$_{7-\delta}$ materials are deduced from the resistivity curves. Itis observed that $\mu_0 H_{irr}$, $\mu_0 H_{C2}$ and $\Delta T_c$ of the film strongly depend on the direction and strength of the field. The coherence length ξ (0) and penetration depth λ (0) values at T = 0 K has been calculated from the irreversibility fields ($\mu_0 H_{irr}$) and upper critical fields ($\mu_0 H_{C2}$) respectively. Based on all the results, the change of the superconducting properties as a function of the magnetic field direction presents the anisotropy of the sample produced.

## 1 Introduction

Since the discovery of superconductivity, many researchers have attempted to improve the mechanical, structural and flux pinning properties of superconducting materials to make them suitable for magnetic field and high temperature applications [1 ‹2 ‹3 ‹4]. The YBa$_2$Cu$_3$O$_{7-\delta}$ (YBCO) have been intensively studied due to their unique properties and their incorporation in superconducting electronics [5]. When the YBCO is exposed to high magnetic fields and temperatures, unusually rapid flux flow starts to appear in the sample, resulting in the energy dissipation and subsequent transition of superconductor to the normal state [6]. In order to overcome these problems appeared, the flux pinning mechanism in the film should be investigated in detail. In the recent works [7, 8, 9], the researchers have extensively analyzed the flux pinning mechanism in the superconducting state to introduce the effective pinning centers such as planar defects, stacking faults, and microdefects, resulting in thermally activated jumps; or hopping of flux lines; or flux bundles over an energy barrier [10, 11]. Over the pinning energy barrier of a structure, the flux line might be thermally activated although the Lorentz force exerted on the flux bundle by the current is smaller than the pinning force [12, 13]. A model, described as thermally activated flux flow (TAFF), works well in the resistivity region near Tc as a result of the zero resistivity (ρ = 0) [14, 15, 16]. Additionally, the width of the superconducting transition depends strongly on the anisotropy associated with the magnetic field direction with regard to the Cu–O planes (the c-axis) in the structure [17, 18, 19]. Flux pinning ability can also be estimated from the flux pinning force density and activation energy values because, as well known, the activation energy (mentioned as the potential barrier height) is generally regarded as a measure of flux pinning strength of a superconductor material [20, 21]. The activation energy is inferred from TAFF theory described with Arrhenius equation $\rho = \rho_0 \exp(-U_0/k_BT)$, which will be dealt with in detail in results and discussion part.

In this study, the role of magnetic field direction on the superconducting of the YBCO thin film is analyzed by the resistivity measurements. It is found that the critical transition temperature decreases with the increase in the applied magnetic field and the change of the applied field direction with regard to the c-axis. Additionally, thermally activated flux flow (TAFF) model is modified to fit the resistivity curves at different applied field strengths up to 14 T. The activation energy, irreversibility field and upper critical field values of the film are determined by the TAFF model.

## 2  EXPERIMENTAL

C-axis oriented epitaxial $YBa_2Cu_3O_{7-\delta}$ films with a thickness of 400 nm and a width of 7.53 μm are deposited by the laser ablation method on the surface (100) of $SrTiO_3$ substrate. The resistance vanished, in zero magnetic field, at $Tc$ = 90 K. Electrodes of measurement are in gold and deposited on the surface of the sample in situ by evaporation. The distance between electrodes of power measurement is 135 μm. Contact resistances were less than 1 $\Omega$. A direct current, perpendicular to the magnetic field, is applied on edges of the sample. The sample central region voltage signal goes through a low-noise transformer of report n = 100, then in a preamplifier of gain equal to 100 and finally in a RC filter. The signal is visualized on a programmable oscilloscope then recorded and analyzed by computer [22].

## 3  Results and discussions

The resistivity measurements of the thin film YBCO as a function of temperature (from 100 K down to 70 K) in the applied magnetic field ranging of 0–14 T are exerted by the standard dc four-probe method at constant driving current of 100 nA. Figures (1) and (2) show the variations of ρ(T) for applied magnetic field parallel to the **C**-axis and **ab** plane, respectively. Resistivity, along the **C**-axis, exhibits a metal-type behavior observed only in well-oxygenated samples [23, 24]. This behavior has sometimes been attributed to a series of short circuits, allowing the current to flow in fact according to the **ab** plans.

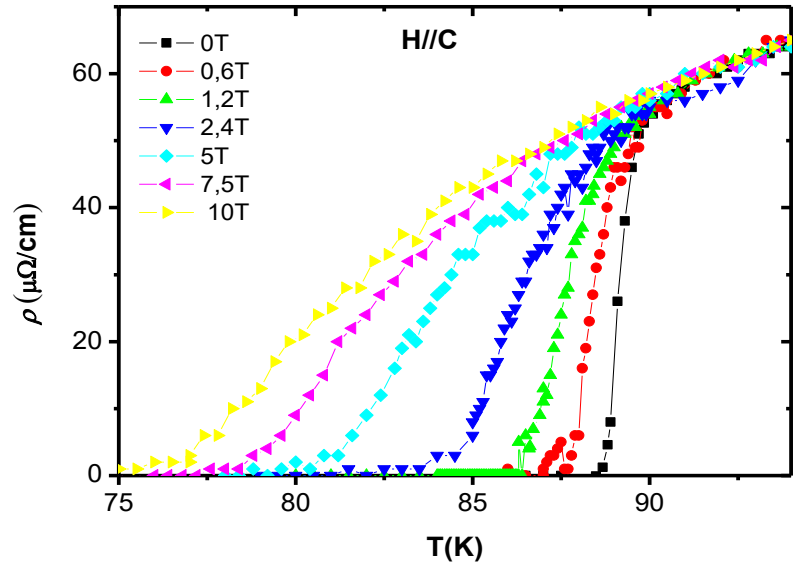

**Figure 1**: The resistivity as a function of temperature for a magnetic field H up to 10T, parallel to the **C**-axis.

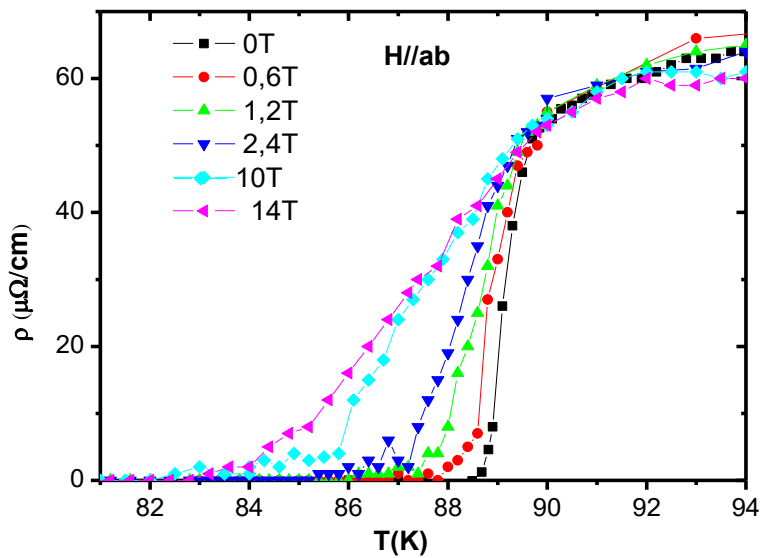

**Figure 2:** The resistivity as a function of temperature for a magnetic field H up to 14T applied parallel to the **ab** plane.

It can be seen in Figure (1) that the resistive transition width increases as the magnetic field increases. In addition, the resistive transition moves to lower temperatures for high magnetic field.

## 3.1 Transition width

Superconducting transition width is defined by $\Delta T_c = T_c^{onset} - T_c^{offset}$, with $T_c^{offset} = 10\%\rho_n$ and $T_c^{onset} = 90\%\rho_n$ ($\rho_n$ resistivity in the normal state). Figure (3) shows the $\Delta T_c$ variation as a function of applied magnetic field, for two angles θ = 0 ° and θ = 90 ° (where θ is the angle between the field and the **C**-axis). Thus, for θ = 90 °, $\Delta T_c$ monotonically increases with magnetic field while for θ = 90 ° an anomalous sharpening of the superconducting transition is observed upon increasing field and takes important values. It can be seen that $\Delta T_c$ increases as the magnetic field increases. The field dependence of $\Delta T_c$ follows a power law as like:

$$\Delta T_c \propto H^n \qquad (1)$$

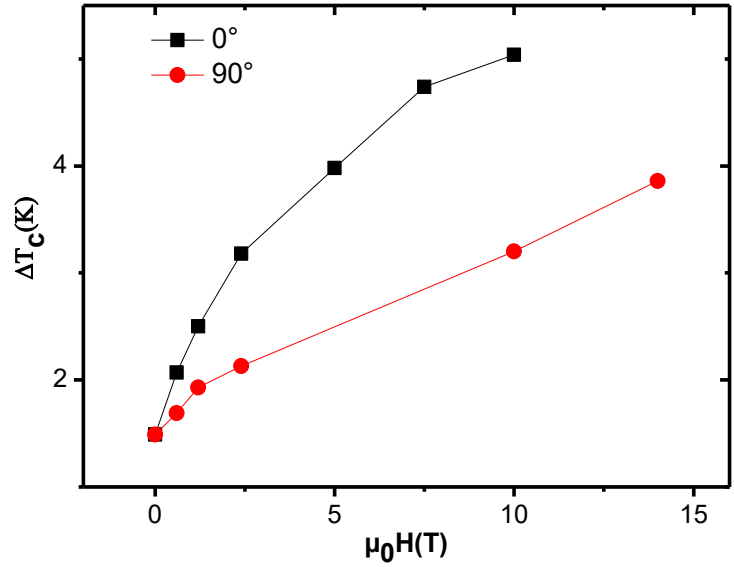

**Figure 3**: Superconducting transition width $\Delta T_c$ versus magnetic field for two orientation angle values.

A good fit of the figure (3), gives a value of **n** = 0.75 for a magnetic field applied parallel to **ab** plane and **n** = 0.57 for one applied parallel to C-axis, the result Alain PAUTRAT [25] YBCO is **n** = 2/3. We note that the transitions have different speeds depending on the direction of application of the field. This difference can be explained by a distribution of defects within the sample [26]. These comparisons show the much narrower thermally activated flux flow (TAFF) region in YBCO crystal.

## 3.2 Irreversibility Field and Upper Critical Field

Estimations of the irreversibility fields ($\mu_0 H_{irr}$) and upper critical fields ($\mu_0 H_{C2}$) at various temperatures were carried out by resistance versus applied field (R–T curves), using the criteria of $0.1R\rho_n$ ($\mu_0 H_{irr}$, T) and $0.9R\rho_n$ ($\mu_0 H_{C2}$, T), respectively [27, 28]. Figures 2a and 2b illustrate the variation of the irreversibility fields and upper critical fields as a function of the temperature. Further, we calculate the $\mu_0 H_{irr}(0)$ and $\mu_0 H_{C2}(0)$ values of the material by using an extrapolation of the curve at absolute zero temperature (T = 0 K) and the results observed are given in Table 1.

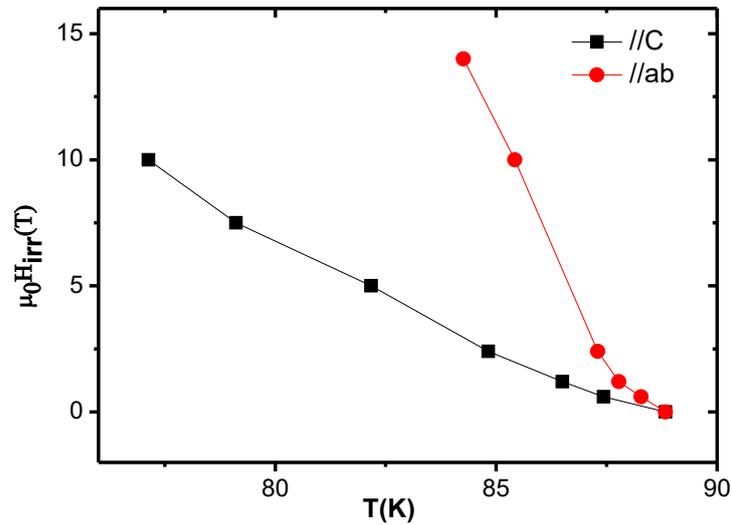

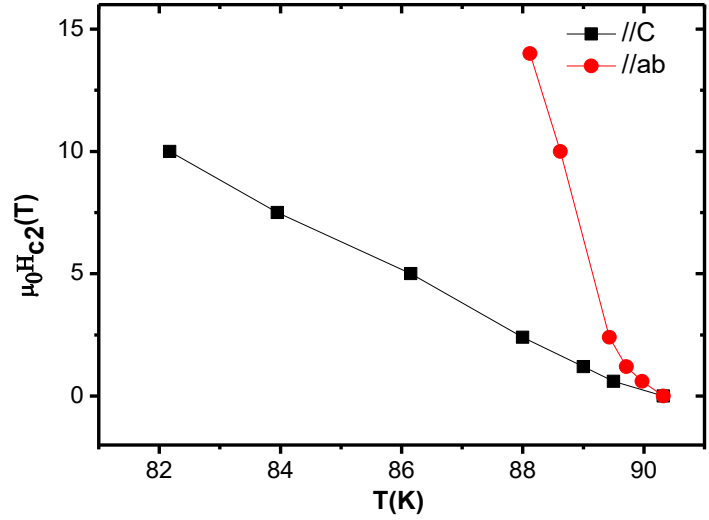

**Figure 4**: **a)** temperature dependences of irreversibility field ($\mu_0 H_{irr}$) and **b)** upper critical field ($\mu_0 H_{C2}$)

The upper critical field $\mu_0 H_{C2}$ is defined as a field where the vortex cores fully fill the material that then becomes normal. The irreversibility line (LI) (or the irreversibility field $\mu_0 H_{irr}$, had associated with the magnetic irreversibility in a measure of the resistivity indicates the vortex stripping or the merging of the flux lines. This line separates the $H/T$ phase diagram into two regions below (LI) there is a critical current while above (LI) $Jc = 0$, excluding any potential application.

The variation of the temperature as a function of the irreversibility field ($\mu_0 H_{irr}$) and upper critical field ($\mu_0 H_{C2}$) determined for the sample is shown in Fig. 2. It is apparent from the figure that the $\mu_0 H_{irr}$ and $\mu_0 H_{C2}$ curves of the sample not only increase with the decrease in the temperature but shift to lower temperatures as the applied magnetic field increases.

It is clearly seen from Fig. 2 that the position of the irreversibility line at θ = 90 ° is higher than those of the θ = 90 °. The decrease of the angle between external magnetic field and c-axis decreases the position of the irreversibility lines, which shows the pinning ability of the sample also decreases with decreasing angle. The change of the irreversibility as a function of the angle indicates the anisotropy of the fabricated film [29].

**Table 1**: Irreversibility field, upper critical field, coherence length and penetration depth values at absolute zero temperature (T = 0 K)

| direction | $\mu_0 H_{C2}(0)(T)$ | $\mu_0 H_{irr}(0)(T)$ | $\xi(0)(Å)$ | $\lambda(0)(Å)$ |
|---|---|---|---|---|
| // ab | 433.54 | 207.84 | 8.71 | 12.59 |
| // C | 79.69 | 54.83 | 20.33 | 24.51 |

By applying the Werthamer-Helfand-Hohenberg (WHH) theory [30], we calculate the values of the critical fields greater than absolute zero (T = 0), for the two orientations of the parallel and perpendicular magnetic field as well as the lines of irreversibility, are grouped in Table1.

$$\mu_0 H_{C2}(0) = -0{,}693\mu_0(dH_{C2}/dT)_{T_c} T_c \quad (1)$$

The $\mu_0 H_{irr}(0)$ values were observed to decrease from 207.84T at θ = 90 °, to 54.83T at θ = 0 °. Likewise, the $\mu_0 H_{C2}(0)$ values were obtained to decrease from 433.54T at θ = 90 °, to 79.69T at θ = 0 °.

The value $\mu_0 H_{C2}/k_B T_C$, calculated from $\mu_0 H_{C2}(0)$, exceeds the Pauli limit (1.84 T / K), indicating the unconventional nature of the superconductivity of our sample.

From Table 1, we see that the field $\mu_0 H_{C2,ab}(0)$, for H // ab, is greater than $\mu_0 H_{C2,C}(0)$, for H // C, this which confirms the anisotropy of the film. This behavior implies a decrease in the pinned capacity.

### 3.3 Coherence Length and Penetration Depth

The coherence length ξ (0) and penetration depth λ (0) values at T = 0 K are described by following equations:

$$\xi(T)^2 = \Phi_0/2\pi\,\mu_0 H_{C2}(T) \text{ and } \lambda(T)^2 = \Phi_0/2\pi\,\mu_0 H_{irr}(T) \quad (2)$$

where Φ0 = 2.07 × 10⁻¹⁵Wb, the coherence length and penetration depth, at absolute zero (T = 0), are given in Table 1. It is apparent from the table that the coherence length ξ at the zero temperature was calculated as 8.71 Å and 20.33 Å for $H_{C2}$ perpendicular and parallel to the C axis, and the penetration depth λ at the zero temperature as 12.59 Å and 24.51 Å for $H_{irr}$ perpendicular and parallel to the C axis, respectively. Hidaka and al. give a value of anisotropy factor of 5.5 in the critical fields [31].

It can be seen that these quantities are larger for an applied magnetic field parallel to the C axis that parallel to the ab plane. This result indicates that the superconducting properties strongly depend on the direction of application of the magnetic field such as anchoring which more important when the applied magnetic field parallel to the **ab** plane. To quantify these differences, the following Ginzburg-Landau anisotropy ratio (AGL) is defined:

$$\gamma = \lambda_c/\lambda_{ab} = \xi_{ab}/\xi_c = H_{c2,ab}/H_{c2,C} \quad (3)$$

which is independent of temperature (in the regime near Tc). We find that γ = 5.44 for our film which in agrees with what found for l'YBa$_2$Cu$_3$O$_{7-\delta}$ [32]. This value of γ implies that H$_{C2}$ has a weak anisotropy. This result indicates that it is a large electron anisotropy in YBCO for its crystallographic anisotropy [31].

## 4 conclusion

The important physical properties of the sample such as $\mu_0 H_{irr}, \mu_0 H_{C2}, \lambda$ and $\xi$ values are inferred from the ρ–T curves under magnetic fields (parallel and perpendicular to **C**-axis) up to 14 T. The results indicate that these properties depend strongly on the magnetic field direction. The variation of the irreversibility field, the upper critical field, the penetration depth and the coherence length depending on the direction of the magnetic field clearly shows the anisotropic of the YBCO thin film. Moreover, $\mu_0 H_{irr}(0)$, and $\mu_0 H_{C2}(0)$ values were determined from the $\mu_0 H_{irr}$ and $\mu_0 H_{C2}$ versus temperature graphs at absolute zero temperature, respectively. The penetration depth and the coherence length are calculated from the irreversibility field and the critical magnetic field greater than T = 0 K, respectively. The change of these parameters as a function of angle gives the information about anisotropy of the film.

## Références

[1]  P. Sarun, S. Vinu, R. Shabna, A. Biju and Syamaprasad, Mater. Res. Bull. 44, 1017–1021, 2009.

[2]  C. Terzioglu, M. Yilmazlar, O. Ozturk and E. Yanmaz, Physica C, 423, 119–126, 2005.

[3]  O. Ozturk, M. Akdogan, H. Aydın, M. Yilmazlar, C. Terzioglu and I. Belenli, Physica B 399, 94, 2007.

[4]  C. Terzioglu, O. Ozturk, A. Kilic, A. Gencer and I. Belenli, Physica C 434, 153, 2006.

[5]  C. Varanasi, J.Burke, H.Wang, J.H.Lee and P.N.Barnes, ,Appl.Phys.Lett.93, 092501, 2012.

[6]  K. Koyama, S. Kanno and S. J. J. Noguchi, Appl. Phys. 29, L53, (1990).

[7]  M. Wakata, S. Takano, F. Munakata and H. C. Yamauchi, 32, 1046, (1992).

[8]  S. Vinu, P. Sarun, S. R. A. Biju and U. Syamaprasad, J. Appl. Phys. 104, 043905, (2008).

[9] E. Asikuzun, O. Ozturk, H. Cetinkara, G. Yildirim, A. Varilci, M. Yılmazlar and C. J. M. Terzioglu, Sci. Mater. Electron, (2011).

[10] X. Wan, Y. Sun, W. Song, L. Jiang, K. Wang and J. Du, Supercond. Sci. Technol. 11, 1079, (1998).

[11] S. Vinu, P. Sarun, R. Shabna, U. Syamaprasad and J. Alloys, Compd. 487, 1, (2009).

[12] C. J. Poole, R. Farach, R. Creswick and R. Prozorov, Superconductivity, 2nd edn. Academic Press, London, (2007).

[13] T. Sheahen, Introduction to High-Temperature Superconductivity,, New York: 1st edn. Springer,, 1994).

[14] A. Malozemoff, T. Worthington, E. Zeldov, N. Yeh, McElfresh and M.W., In: Fukuyama, H., Maekawa, S., Malozemoff,A.P. (eds.) Strong Correlation and Superconductivity. Springer Series, Berlin: in Sol. State Sci., vol. 89. Springer,, 1989).

[15] R. Ma, W. Song, X. Zhu, Zhang, L. Liu, S. Fang, J. Du, J. Sun, Y. Li, C. Yu, Z. Feng and P. Y. Zhang, Physica C 405, 34, (2004).

[16] R. Griessen, Phys. Rev. Lett. 64, 1674, (1990).

[17] M. Charalambous, J. Chaussy and P. Lejay, Phys. Rev. B 45, 5091, 1992.

[18] Yildirim.G, M. Dogruer, O. Ozturk, A. Varilci, T. C. and Y. Zalaoglu, J. Supercond. Nov. Magn., (2011)..

[19] A. Salem, G. Jakob and H. Adrian, Physica C 402, 354, (2004).

[20] M. Pu, W. Song, B. Zhao, X. Wu, Y. Sun, J. Du and J. Fang, Fang, J.: Physica C 361, 181, (2001).

[21] G. Yildirim, M. Akdogan, S. Altintas, M. Erdem, C. Terzioglu and A. Varilci, Physica B 406, (1853), 2011.

[22] R. F. Voss and J. Clarke, Phys. Rev. B 13, 556, 1976.

[23] L. Forro, C. Ayache, J. Henry and J. Rossat-Mignod, Phys. Scr. 41,365., 1990.

[24] T. Ito, H. Takagi, S. Ishibashi, T. Ido and S. Uchida, Nature 350, 596, (1991)..

[25] A. PAUTRAT, STRUCTURE ET DYNAMIQUE DU RESEAU DE VORTEX D'UN SUPRACONDUCTEUR DANS LE CADRE D'UN ANCRAGE DE SURFACE, these,, 2000..

[26] F. HARDY, Etude du composé ferromagnétique supraconducteur URhGe, THESE,, 2004.

[27] M.S.Ososky, R.J.Soulen, S.A.Wolf, J.M.Broto, J.M.Rakoto, J.C.Ousset, G.Coffe, S. Askenazy, P.Pari, I.Bozovic, J.N.Eckstein and G.F.Virshup, Phys.Rev.Lett.71, 1993, p. 2315.

[28] M.Erdem, O.Ozturk, E.Yucel, S.P.Altintas, A.Varilci, C.Terzioglu and I.Belenli, Physica B 406, 2011, p. 705.

[29] G. Yildirim, M.Akdogan, S.P.Altintas, M.Erdem, C.Terzioglu and A.Varilci, "Investigation of the magnetic field angle dependence of resistance, irreversibility field, upper critical field and critical current density in DC sputtered Bi-2223 thin film," in *Physica B406*, 2011, p. 1853–1857.

[30] N. R. Werthamer, E. Helfand and P. C. Hohenberg, Temperature and Purity Dependence of the Superconducting Critical Field, H„. III. Electron Spin and Spin-Orbit Effects, PHYSICAL REVIEW, 1 4 7, 1, p295, 1966..

[31] Y. Hidaka, Y. Enomoto, M. Suzuki, M. Oda, A. Katsui and T. Murakami, Anisotropy of the Upper Critical Magnetic Field in Single Crystal YBa2Cu3O7+y, The Japan Society of Applied Physics,, 1987..

[32] E. Carreño-Morelli, Supraconductivité HTc et Flux magnétiques, 1999. .